\documentclass[letterpaper]{sig-alternate-2013}

\permission{}
\conferenceinfo{\textbf{In Proceedings of the 25th International WWW Conference, 2016}}{\textbf{PLEASE CITE ONLY CONFERENCE PROCEEDINGS, NOT THIS PREPRINT}}
\copyrightetc{ACM \the\acmcopyr}
\crdata{978-1-4503-4143-1/16/04. \\
http://dx.doi.org/10.1145/2872427.2883019}

\usepackage{url}
\usepackage{stfloats}
\usepackage{color}
\usepackage{subfigure}

\newcommand{\mean}[1]{\left\langle #1 \right\rangle}

\newcommand{\mst}[1]{\textbf{\color{red}/* #1 (mst) */}}

\begin{document}

\title{The QWERTY Effect on the Web}
\subtitle{How Typing Shapes the Meaning of Words in Online Human-Computer Interaction}

\numberofauthors{2}
\author{
\alignauthor
        David Garcia\\
       \affaddr{ETH Zurich}\\
       \affaddr{Zurich, Switzerland}\\
       \email{dgarcia@ethz.ch}
\alignauthor
        Markus Strohmaier\\
       \affaddr{GESIS - Leibniz Institute for the Social Sciences and University of Koblenz-Landau}\\
       \affaddr{Cologne, Germany}\\
       \email{markus.strohmaier@gesis.org}
}
\maketitle
\begin{abstract}

The QWERTY effect postulates that the keyboard layout influences word meanings
by linking positivity to the use of the right hand and negativity to the use
of the left hand. For example, previous research has established that words
with more right hand letters are rated more positively than words with more
left hand letters by human subjects in small scale experiments. In this paper,
we perform large scale investigations of the QWERTY effect on the web. Using
data from eleven web platforms related to products, movies, books, and videos,
we conduct observational tests whether a hand-meaning relationship can be
found in text interpretations by web users. Furthermore, we
investigate whether writing text on the web exhibits the QWERTY
effect as well, by analyzing the relationship between the text of online
reviews and their star ratings in four additional datasets. Overall, we find
robust evidence for the QWERTY effect both at the point of text interpretation
(decoding) and at the point of text creation (encoding). We also find under
which conditions the effect might not hold. Our findings have implications for
any algorithmic method aiming to evaluate the meaning of words on the web,
including for example semantic or sentiment analysis, and show the existence
of "dactilar onomatopoeias" that shape the dynamics of word-meaning
associations. To the best of our knowledge, this is the first work to reveal
the extent to which the QWERTY effect exists in large scale human-computer
interaction on the web.

\end{abstract}



\section{Introduction}

\emph{What were the body organs that you used the last time that you
communicated with someone?} If we asked this strange question 200 years ago,
the most likely answer would be: \emph{"my vocal cords"}. With the rise of the
web and online communication, we are increasingly more likely to get another
answer today: \emph{"my fingers"}.  With more than 200 Billion emails sent per
day \cite{Radicati2015} and textual communication being more frequent than
voice calls in mobile phone usage \cite{Smith2011}, there is little doubt that
the human-computer interface, in  particular fingers and keyboards, plays an
important role in online communication and the web in general.

\textbf{The QWERTY effect} postulates that the keyboard layout influences word
meanings by linking positivity to the use of the right hand and negativity to
the use of the left hand. In \cite{Jasmin2012} for example, Jasmin and
Casasanto show that the QWERTY effect\footnote{Not to be confused with the
QWERTY network effect in economics to explain its majority use.} manifests in offline small scale
experiments, as human subjects rated those words more positively that were typed with more
letters from the right side of the keyboard than words typed with more letters
from the left \cite{Jasmin2012,Casasanto2014}. General lateralization effects
explain the existence of the QWERTY effect, including the positive connotation
of the \emph{right} \cite{Casasanto2014b} and the positive effect of fluency
that right-handed typers can experience when using the right hand more often.

\begin{figure*}[t] \centering
\includegraphics[width=0.95\textwidth]{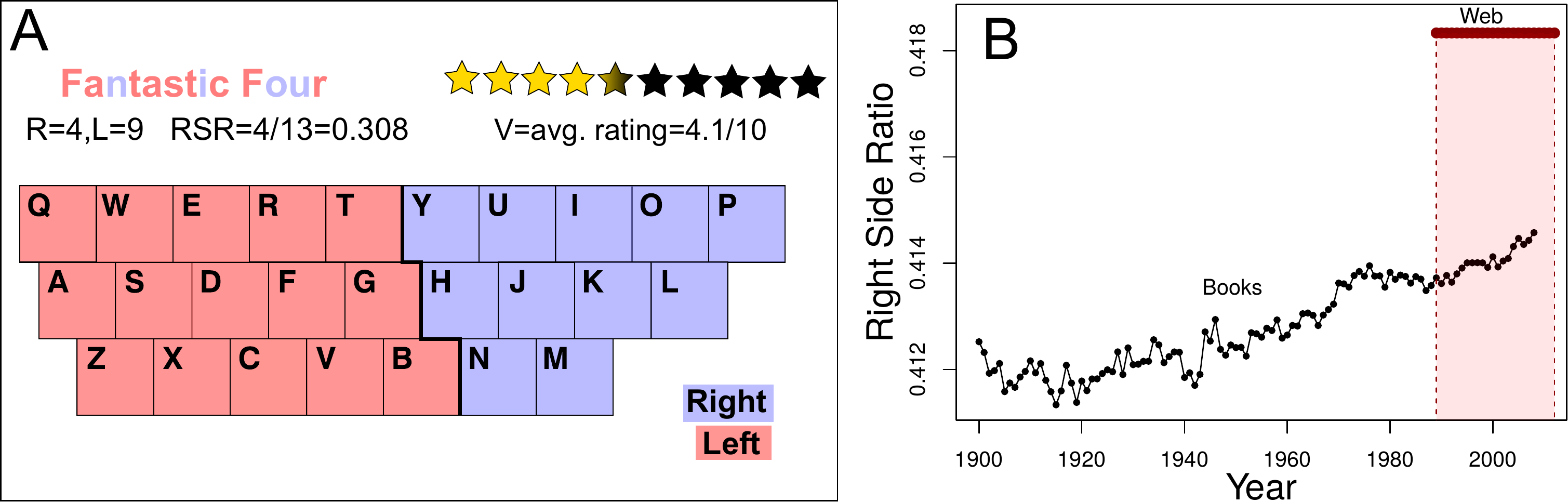}\\ \caption{ A)
Hand usage in the QWERTY layout and an example of the Right Side Ratio (RSR) and
valence $V$ using the movie 'Fantastic Four' on IMDB. B) RSR on the web and books. Black
dots show the yearly RSR of Google books \cite{Lin2012} published since 1900.
The red bar shows the empirical value of RSR in an English-speaking web text
corpus \cite{Jakubivcek2013}. The ratio increases with time and has a higher
value on the web. \label{fig:BooksTS}} \end{figure*}

We measure the tendency to use right side letters through the Right Side Ratio
(RSR), calculated as the amount of right side letters divided by the sum of
right and left side letters in a text. Figure \ref{fig:BooksTS}A shows the
general pattern of hand usage in the QWERTY layout and an example of how the
RSR is calculated.  Figure \ref{fig:BooksTS}B shows the RSR for English books
of the Google Books corpus \cite{Lin2012} published each available year since
1900, illustrating a general upwards trend of this ratio that specially speeds
up in the early 1990s. The vertical red bar shows the RSR computed for a
corpus of English text on the web \cite{Jakubivcek2013}, displaying a higher
tendency of right side letters on the web as compared to books. What remains
unknown from this preliminary analysis is whether the increasing prevalence of
words with a higher RSR on the web translates into evidence for the QWERTY
effect.

We thus set out to study \emph{whether evidence for (or against) the QWERTY
effect can be found on the web}. In other words, we  want to test
the hypothesis that textual web content with more right side letters is
evaluated more positively by users. In addition, we also want to study
boundary conditions of the QWERTY effect on the web, i.e. the extent to which
it can be observed, and its limitations.

Building on existing methods used to analyze small scale experimental data
\cite{Casasanto2014}, we perform a large scale confirmatory analysis to test
if the QWERTY effect manifests on the web. We test if the QWERTY effect is
present when \emph{(i) decoding text}, i.e. when evaluating items with names
or titles (e.g. IMDB movie titles). For that, we gather eleven datasets of
online media, recording how content is evaluated via votes, likes, or ratings.
Using that data, we explore the existence of a hand-meaning relationship
through a series of statistical tests, including permutation tests, robust
regression, and control for linguistic and contextual factors that could
influence the effect. We also test the QWERTY effect when \emph{(ii) encoding
text}, i.e. when online users write text to express certain meaning (e.g.
Amazon reviews). For that, we analyze four additional datasets of online product reviews
that are accompanied by a summary of their evaluative meaning as a star
rating. On these datasets, we test the existence of a relationship between
star ratings of a review and the amounts of right and left side letters of its
text.

We find mostly consistent, significant evidence for the existence of the
QWERTY effect across 15 web datasets, both on an encoding and decoding level.
More specifically, we find that the ratio of right side letters has a
normalized coefficient between 1 and 3\% of the standard deviation of average
evaluations given by users, and the amount of left-hand letters in a review
are associated with decreases of up to 17\% of a star per 100 letters (more
details in Section \ref{sec:encoding}). These apparently small effect sizes
are characteristic of psycholinguistic studies, where word frequencies usually
variate at the 1\% level \cite{Golder2011,Kramer2014}.  We analyze the
interaction between these effects and linguistic and contextual variables,
finding robust evidence of the effect against a range of plausible confounds.
Our analysis also reveals some interesting limitations of the QWERTY effect,
which can weaken for popular web content or can even be reversed in the
language of very particular contexts. The results reported in this paper are
limited to English, and our models are only testable when explicit positive
and negative ratings are present.

Our contributions are two-fold: (i) To the best of our knowledge, this is the
first large scale attempt to study the existence of the QWERTY effect on the
web. (ii) in addition to decoding studies which have been performed by
previous research in small scale, offline experiments, we are also testing if the QWERTY effect exists when
encoding text. Overall, our contribution lies in the execution of a confirmatory study to
test the extent of the QWERTY effect on the web, and its conditionants.

\section{Research background}

\textbf{The QWERTY effect:} The QWERTY effect was first reported in secondary
analyses of word rating experiments \cite{Jasmin2012}, testing the
relationship between the valence expressed in a word and its difference
between the amounts of right and left side letters in the word. The
explanation of the effect existing due to the keyboard is specially
consistent, as the effect is stronger for words created after the
popularization of the keyboard and even for pseudowords without explicit
meaning \cite{Casasanto2014}. Besides this robust evidence in experimental
tests, the state-of-research on this effect is limited in two ways: First,
previous research focused on word interpretation (decoding) and did not test
the existence of the effect when writing text. Second, results so far only
show the existence of the effect in a precise controlled scenario, leaving
open whether it appears in practice, for example if products with \emph{the
"right" name} have an advantage \cite{Jasmin2012}. We aim at contributing to
our knowledge of the QWERTY effect in a way that complements these two
limitations, through a large scale observational analysis of the effect on the
web that includes both encoding and decoding scenarios.

The fact that we cannot generally perceive the relationship between hand use
and meaning suggests that the effect must be very subtle, difficult to measure
with limited data, and easy to distort through experimental biases. One of our
aims is to test if the effect exists when encoding, which is difficult to
measure in the laboratory, since it would require participants to produce
large amounts of text. To date, the only observational evidence of the effect
during encoding focuses on baby names, testing if there is a bias towards
right side letters when naming newborns. Results for this case are mixed,
while there is a general increase of right side letter use in US baby names
\cite{Casasanto2014}, a precise test of a correlation between right side
letter use and baby name popularity yields negative results
\cite{Thogmartin2013}.

\textbf{Explanations for the QWERTY effect:} \emph{How can typing influence
word meaning?} Two possible mechanisms can be identified in the literature.
First, the fluency experienced when using the right hand and its associated
pleasantness can drive right-hand typers to use right keys when expressing
positivity. Since the majority of society is right-handed and the meaning of
words is normative, a general association between right-handed words and
positive meanings can emerge \cite{Jasmin2012}. Second, the QWERTY effect
could be explained by general lateralization: individuals have implicit biases
to associate right to good and left to bad, even across cultures
\cite{Casasanto2014b}. In fact, lateralization might be biologically grounded,
as experiments with newborn chicken show faster learning rates when positive
stimuli are located on the right \cite{Vallortigara1996}. Furthermore,
smartphone usage is linked to increased finger sensitivity at the neural level
\cite{Gindrat2015}, motivating the explanation that the QWERTY effect could be
built on the combination of biological and technological factors.

\textbf{Communication of emotion:} Among the possible ways to measure meaning,
our research question deals with evaluation, formalized for example in the
semantic differential \cite{Osgood1964}, and commonly named "\emph{valence}"
in emotion research \cite{Russell1977,Scherer2005}. The valence associated to
a word or a text measures the degree of pleasure or displeasure, the degree of
positivity or negativity that the word or text means. The communication of
emotions is integrated in a wider perspective in the \emph{hyperlens model} of
emotions, and its composed of two components: (i) encoding, in which the
transmitter turns emotional meaning into signs, and (ii) decoding, when the
receiver transforms those signs back into meanings \cite{Kappas2013}. In this
model, encoding and decoding happen in the presence of individual feedback and
social and cultural contexts, considering emotions as interpersonal (social)
phenomena that are embodied and not only present in the mind. The case of the
QWERTY effect is specially interesting for the concept of embodied emotions,
since it proposes a fundamental change in emotional communication induced by a
physical constraint, the keyboard layout, and not by any cognitive phenomenon.

\textbf{Expression of sentiment on the web:} Online human-computer interaction
has an important transformative power in society \cite{Dimaggio2001}, changing
our attitudes \cite{Sassenberg2003}, and leaving traces that allow us to
understand human behavior like never before \cite{Golder2014}. Through the
digital traces of the web we can learn social dynamics of emotions, for
example the daily patterns of emotional expression \cite{Golder2011}, the
emergence of collective emotions \cite{Garas2012}, or the spreading of
emotional messages \cite{Alvarez2015}. Fundamental properties of
language can also be analyzed this way, for example the existence of positive
biases across languages \cite{Garcia2012,Dodds2015}, the assortativity of
subjective expression \cite{Bollen2011}, and the community components of
language \cite{Danescu2013,Bryden2013}. Analyzing the content of online data
has promising applications through sentiment analysis, from monitoring
reactions to politically relevant events \cite{Gonzalez2012} to financial
trading based on emotional expression \cite{Bollen2011b,Garcia2015}. Thus,
understanding how language might be influenced by technology is not only
interesting as a fundamental research question, but has implications for the
application of online data analysis.


\section{Decoding study}
Our aim is to test the QWERTY effect in a wide variety of
settings, including possible variations of language style and communication
mechanisms.
\paragraph{Datasets} To test the effect, we need datasets that contain items (products,
movies, videos, etc) that are rated in an evaluative manner, with both
positive and negative feedback that can be aggregated to a valence score. If
only positive feedback or views are present, the question would become about
popularity rather than about evaluation. In addition, we need items to be
named with a character string that is used to refer to them. For example,
product names or movie titles fall on this class, while summaries or question
formulations do not constitute symbols that could be affected by the QWERTY
effect. Since the scope of this research is to try the QWERTY effect as it has
been formulated for the English keyboard layout, we will focus on English-speaking online communities and filter out non-English data in case of its
existence. A summary of these datasets is shown on Table \ref{tab:data1}.

\begin{table*}[bp]
\begin{center}
    \begin{tabular}{c|c|c|c|c|c|c}
Dataset &   N elements & $\mean{V}$ & scale & $\mean{RSR}$ & additional controls & source \\ \hline

Amazon & 4,257,624 & 3.86 & 1-5 & 0.4176 & sales rank, price, $N_r$ & \cite{Mcauley2015,Mcauley2015b} \\
Yelp & 56,103 & 3.66 & 1-5 & 0.4056 & $N_r$ & Yelp Challenge Dataset$^2$ \\
Epinions & 223,880 & 3.89 & 1-5 &  0.4174 &  $N_r$ & \cite{Tanase2013} \\
Dooyoo & 112,698 & 3.89 & 1-5 &  0.4184 & $N_r$ & \cite{Tanase2013} \\
IMDB & 327,608 & 6.30 & 1-10 & 0.425  &  year, isInEnglish, $N_r$ & OMDB\cite{Fritz2015} \\
Rotten Tomatoes & 80,756 & 3.04 & 1-5 & 0.4233 & $N_r$ &  OMDB\cite{Fritz2015} \\
MovieLens & 29,505 & 3.11 & 1-5  & 0.4245 &  $N_r$ & University of Minnesota Dataset$^3$\\ 
BookCrossing & 149,804 & 7.42 & 1-10  & 0.4164 & $N_r$ & \cite{Ziegler2005} \\
Youtube & 3,292,153 & 0.94 & 0/1 & 0.4294 & views, comments, date, $N_r$ & \cite{Abisheva2014} \\
Redtube & 351,677 & 0.70 & 0/1 & 0.4225 & date & new \\
Pornhub & 333,967 & 0.83 & 0/1 & 0.4264  & views, date, $N_r$ & new\\
	  \end{tabular}
\end{center}
  \caption{Datasets used for the Decoding Study. $V$ refers to valence, RSR refers to the Right Side Ratio.
All datasets included linguistic controls of amount of letters, words, average letter frequency and average word frequency, plus the contextual controls listed on the table.  \label{tab:data1}}
\end{table*}

\textbf{Product and business ratings:} The first class of datasets we
include are general product and business ratings from Amazon.com, Yelp.com,
Epinions.com and Dooyoo.co.uk. We use the deduplicated Amazon dataset from
\cite{Mcauley2015,Mcauley2015b}, focusing only on products with complete
metadata that includes price and sales rank. This way, we count with more than
4 million rated products. For the case of Yelp, we analyze the Yelp challenge
dataset\footnote{\url{http://www.yelp.com/dataset_challenge}}, filtering out business from non English-speaking
locations. As a result, we count with the ratings and names for more than 55
thousand businesses.  Data on Epinions and Dooyoo was provided as part of
social recommender systems research \cite{Tanase2013}, including more than 200
thousand and 100 thousand rated products in Epinions and Dooyoo respectively.

\textbf{Movie and book ratings:}  The second class of datasets we include is
based on reviews of cultural items like movies or books. First, we downloaded
the datasets of the Open Movie DataBase \cite{Fritz2015}, which is based on
data publicly displayed on IMDB.com and RottenTomatoes.com and  has been used
in previous research on movies \cite{Garcia2014}. For IMDB, we count with the
average rating provided by users on a 1 to 10 star scale, and with additional
information about the movie including year of release and languages for more
than 300 thousand movies. The case of Rotten Tomatoes is smaller but includes
more than 80 thousand movies rated by the users of the platform (we discard
professional critic data).  The third movie dataset we include comes from the
MovieLens platform and is publicly available for research
\footnote{\url{http://grouplens.org/datasets/movielens/}}, including ratings in a 1
to 5 scale for more than 29 thousand movies. We complete this group of
datasets with the BookCrossing dataset \cite{Ziegler2005} that includes
ratings on a 1 to 10 scale for almost 150 thousand books.

\textbf{Video ratings:} The third class of data we analyze are video sharing
communities, in which users rate videos with up and down votes, or "likes" and
"dislikes". We start from one of the largest video sharing communities,
Youtube, using a dataset of a large scale crawl of Youtube videos
\cite{Abisheva2014}. To filter out non-English videos, we applied language
detection \cite{Danilak2015} and only included videos with a title identified
as English. This includes more than 3 million videos in our analysis with
contextual data that contains the upload date and the amounts of views and
comments of each video. We replicate this dataset with our own crawl of two
adult video communities that have a design very similar to Youtube. Using the
sitemaps described in the robots.txt file of these sites, we retrieve the
title and upvote ratios, as well as some additional data as in Youtube,
including views and upload date. This adds two more video datasets with more
than 300 thousand videos each.

\paragraph{Methods}

\textbf{Variables:} Our approach builds on the methods of \cite{Jasmin2012},
measuring three relevant types of variables for each item that is evaluated by
an online community. First, we compute the Right Side Ratio ($RSR$) of the
item's name as $RSR=R/(R+L)$, where $R$ is the amount of instances of right
side letters of the QWERTY keyboard and $L$ of left side letters.  Second, we
estimate the valence $V$ of the evaluation of the item as provided by the
community, by averaging the votes assigned to the item. The way we calculate
$V$ depends on the community, for example for IMDB movies we compute the
average star rating (as shown in the example of Figure \ref{fig:BooksTS}A),
while for videos we take the ratio of likes over the sum of likes and
dislikes. Third, we compute a set of controls to include in our analysis, in
order to test possible factors that can increase or diminish  the QWERTY
effect. We distinguish two types of controls: i) linguistic controls that can
be computed for all datasets, including the amount of letters and words in the
name and the average letter and word frequency as computed in the Google books
dataset \cite{Lin2012}, and ii) contextual controls that include community-
dependent observable variables such as the amount of views of a video, the
year of a movie, or the price of a product.

\textbf{Statistical tests:} In our analysis, the valence of an item is the
dependent variable that is hypothesized to increase with the RSR. Thus, our
first test  is based on a regression model
\begin{equation}
V= a + b*RSR + \epsilon \label{eq:1}
\end{equation}
in which the QWERTY effect states the alternative hypothesis $b>0$
versus the null hypothesis of $b=0$. This way the  \emph{right side
coefficient} $b$ measures the tendency of the valence of items to grow with
the $RSR$ of their names.  On top of this model, we perform a series of
statistical tests. First, we perform traditional Ordinary Least Squares (OLS)
and test the significance of the estimate $\hat{b}$ with the standard Student
t-test. This test has the advantage of being general and simple, but implies a
series of assumptions: normality of $\epsilon$, sample independence, and
homoskedasticity. We provide additional tests that relax these assumptions and
add robustness to our statistical analysis. We perform a permutation test
\cite{Good2006}, in which we compute the same estimate $\hat{b}$ over $10000$
randomized versions of the data where the names of products have been
reshuffled.   Furthermore, we perform bootstrap tests over the OLS estimate,
using $10000$ bootstrap samples with replacement \cite{Davison1997}.
Estimating $\hat{b}$ over these samples, we compute a bootstrap confidence
interval to further test the hypothesis of a positive right side coefficient.
To cope with heteroskedasticity, we apply an MM-type robust regressor
\cite{Yohai1987}, which controls the leverage of outliers and
provides residuals closer to normality than OLS. In addition, we compute
Spearman's correlation between $V$ and $RSR$ on a bootstrap test, further
testing if a positive relation between the $RSR$ and $V$ exists when outliers
are linearized into ranks.

\begin{figure*}[ht] \centering
\includegraphics[width=0.98\textwidth]{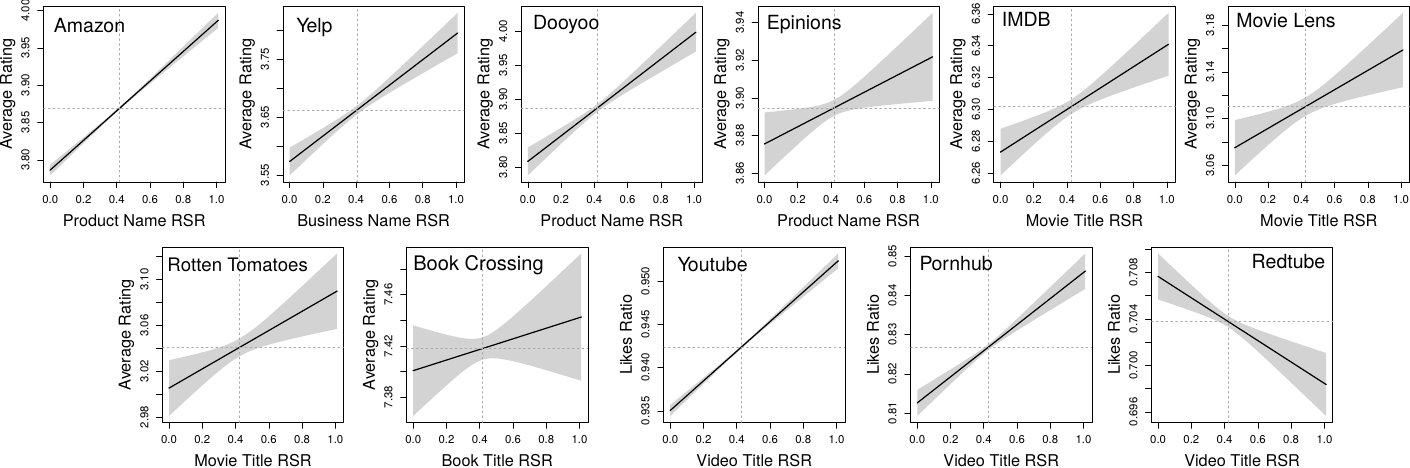}\\
\includegraphics[width=0.98\textwidth]{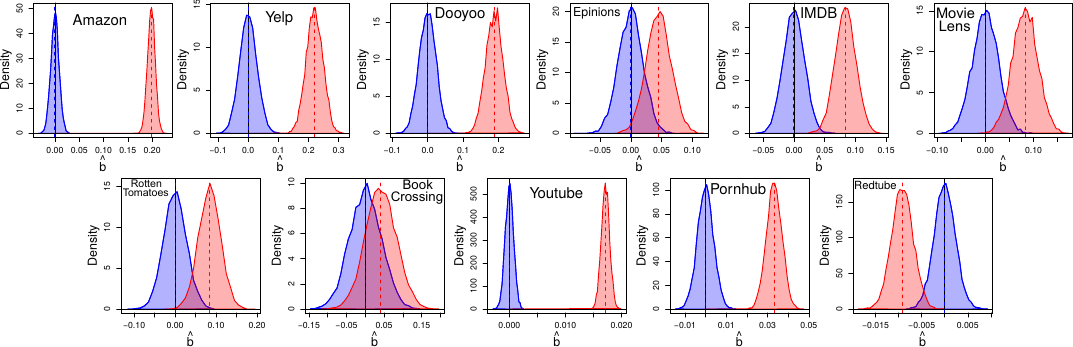}\\ \caption{Top:
Estimates of the QWERTY effect when decoding text based on 11 web datasets. The solid line shows
the estimated value of valence V as a function of the Right Side Ratio (RSR) of the evaluated content,
the gray area shows the standard error around the estimate in the linear
model.  Bottom: Estimates of the right side coefficient $\hat{b}$. The figure shows density functions of bootstrap estimates of the QWERTY effect
(red) and estimates under permutation (blue). Vertical dashed lines show the
mean estimates. The hypothesis of the QWERTY effect is generally confirmed,
with the exception of the cases of BookCrossing and Redtube.  The estimates
under permutation are concentrated around 0 and do not reach the mean point
estimates. The distributions of $\hat{b}$ in bootstrap tests are significantly
above 0 in the same cases as in the Ordinary Least Squares (OLS) estimate. \label{fig:Decoding}}
\end{figure*}

\textbf{Controlling for confounds:} The effect size suggested by previous
findings is very small but very general, motivating the inclusion of
linguistic controls to discard possible confounds with  frequency or sample
size variables \cite{Casasanto2014}. To test if our findings can be explained
due to confounds or sample biases, we perform an additional test with linear
controls. The test works as follows: first, we fit a model of the form
\begin{equation}
V = \sum_i  c_i * X_i + V_r
\end{equation}
where $X_i$ are a set of control variables. $V_r$ is the residualized valence,
i.e. the remaining valence that cannot be explained by the controls and
constitutes the error term of the regression. Then we fit the a model
\begin{equation}
V_r = a' + b'*RSR +\epsilon'
\end{equation}
in order to test the hypothesis $b'>0$. The control variables depend on the
dataset under study, for example for the case of Youtube: $X = [N_w, N_l, A_w,
A_l, N_r, N_v, N_c, t]$. The linguistic controls are the amount of words
$N_w$, the amount of letters $N_l$, the average word frequency $A_w$, and the
average letter frequency $A_l$. The contextual controls given by Youtube are
the amounts of votes $N_r$, views $N_v$, and comments $N_c$, and the timestamp
of creation of the video $t$. The linguistic controls are common for all our
tests, and the contextual controls of each dataset are listed on Table
\ref{tab:data1}. Besides the  residualized regression, we leverage the large
size of the Amazon and Youtube datasets to test the general trend of the
QWERTY effect, as estimated through the  right side coefficient. Using each
control variable, we stratify the data into deciles according to each variable
and make parallel fits over each stratum. We then analyze  how the right side
coefficient as estimated in Equation \ref{eq:1} might depend on each control
variable, to explore the generality of the effect and evaluate the conditions
that might weaken or strengthen it.

\subsection{The QWERTY effect in decoding text}

The hypothesis $b>0$ was confirmed in 9 out of the 11 datasets in the t-test
of the OLS estimate, as shown on the regression functions of the top panel of
Figure \ref{fig:Decoding}. The results of bootstrap estimates of the right
side coefficient and permutation tests are shown in the bottom panel of Figure
\ref{fig:Decoding}, illustrating how the effects revealed by the model vanish
under permutation. This shows that our methods do not introduce false
positives, as permuted datasets clearly show nonsignificant results with means
around $0$. All these 9 cases reach the significant conclusions in the
bootstrap, permutation and Spearman tests. As reported in Table
\ref{tab:Decoding}, these cases reach the maximum level of statistical
significance ($p<0.05$) reachable in $10000$ samples of the bootstrap,
permutation, and Spearman tests.

Our additional robustness test includes an MM-type estimator that lowers the
influence of outliers.  All the cases passing the above tests also pass this
test, with the exception of Epinions, which shows no significant results when
outliers are corrected (see Table \ref{tab:Decoding}). It is worth noting that
in some cases, the size estimate of the right side coefficient $\hat{b}$  was
slightly moderated in robust regression, but it never changes sign.

\begin{figure}[t]
\centering
\includegraphics[width=0.48\textwidth]{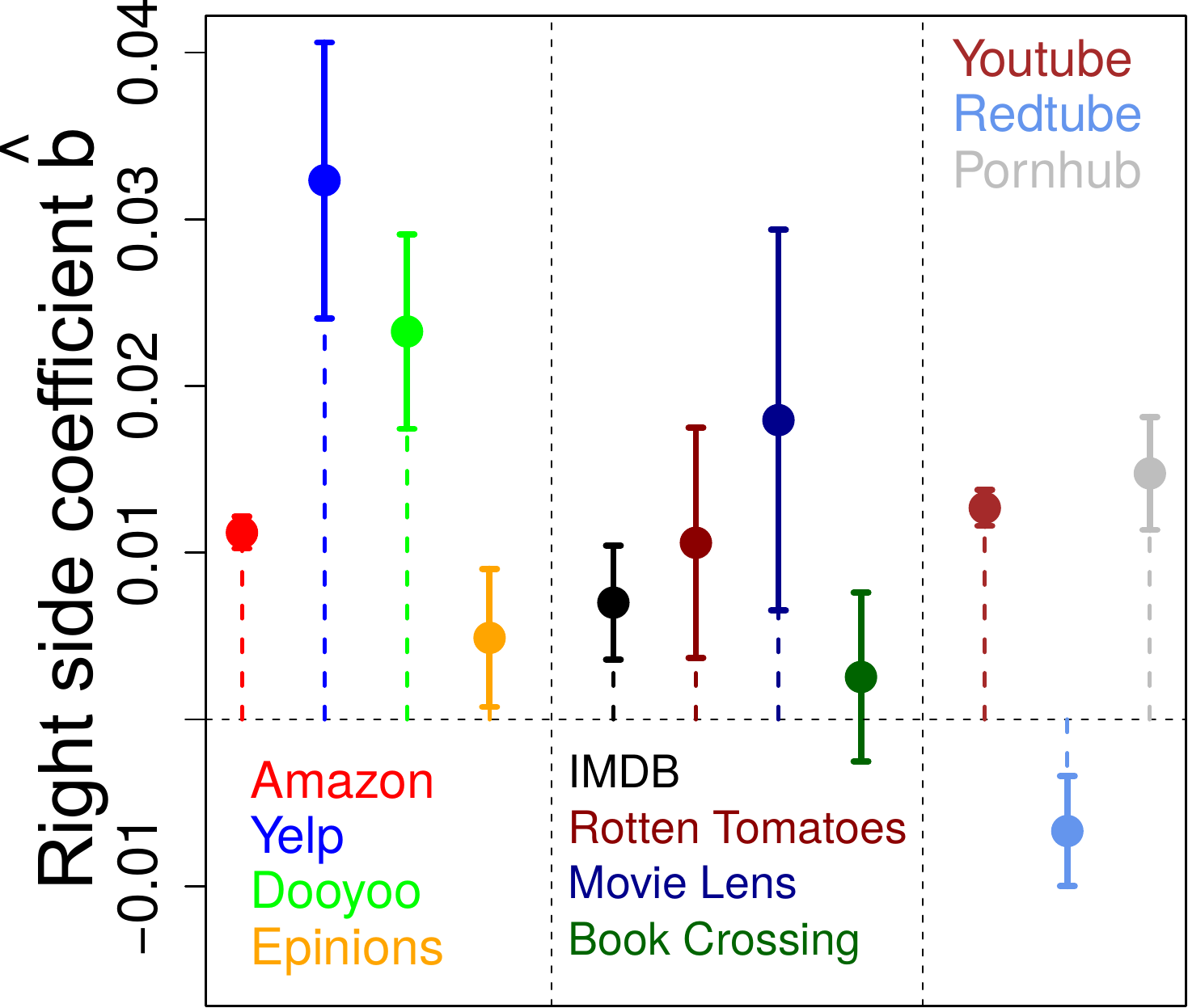}\\
\caption{Summary of the right side coefficient $\hat{b}$ in decoding text. Dots show point estimates of the normalized linear model, error bars show 95\% confidence intervals.  The estimate of the right side coefficient is generally positive and significant, with values up to 3\% of the standard deviation of $V$. \label{fig:Summary}}
\end{figure}

To compare effect sizes, we performed a normalized version of the OLS test, in
which we computed the Z-score of each variable. In this comparison, the right
side coefficient  is measured as a ratio of standard deviations, rather than
in the natural units of each case. Figure \ref{fig:Summary} shows the point
estimates of this normalized right side coefficient $\hat{b}$  and their
$95\%$ confidence intervals. The effect size is the largest in Yelp, followed
by Dooyoo, and the estimate is very small for the case of Epinions, in line
with the non-significant robust regression result. It is interesting to point out
that one of the adult video communities, Redtube, displays a reversed effect
with negative right side coefficient (all regression results had $p<0.001$ and
negative point estimates, permutation, bootstrap and Spearman one-tailed
tests gave $p=1$ against $b>0$). This suggests that the hand-meaning relation
does not need to manifest in the same way in all contexts. We provide more
details on possible reasons for this phenomenon in the discussion.

\begin{table*}[bp]
\begin{center}
    \begin{tabular}{c|cc|cc|cc|c|c|c}

Dataset 		& \multicolumn{2}{|c|}{OLS $\hat{b}$}  & \multicolumn{2}{|c|}{MM $\hat{b}$}  & \multicolumn{2}{|c|}{Residualized  $\hat{b}$}  & Bootstrap & Permutation & Spearman \\
\hline
Amazon  		& $0.1984$ & $p<0.001$  & $0.1384$ & $p<0.001$  & $0.0348$ & $p<0.001$ & $p<0.05$ & $p<0.05$ & $p<0.05$ \\
Youtube 		& $0.0171 $ & $ p<0.001$  & $0.0007 $ & $ p<0.001$  & $0.0109$ & $p<0.001$ & $p<0.05$ & $p<0.05$ & $p<0.05$ \\
IMDB 			& $0.0670 $ & $ p<0.001$  & $0.0618 $ & $ p<0.001$  & $0.0887$ & $p<0.001$ & $p<0.05$ & $p<0.05$ & $p<0.05$ \\
Yelp 			& $0.2192 $ & $ p<0.001$  & $0.2303 $ & $ p<0.001$  & $0.2964$ & $p<0.001$ & $p<0.05$ & $p<0.05$ & $p<0.05$ \\
MovieLens		& $0.0826 $ & $ p<0.001$  & $0.0715 $ & $ p<0.01$   & $0.1128$ & $p<0.001$ & $p<0.05$ & $p<0.05$ & $p<0.05$ \\
BookCrossing	& $0.0414 $ & $ p>0.1$  	& $0.0516 $ & $ p>0.1$    & $-0.0408$ & $p>0.1$  & $p>0.1$ & $p>0.1$ & $p>0.1$ \\	
Epinions		& $0.0458 $ & $ p<0.05$  &  $0.0182 $ & $ p>0.1$    & $0.0360 $ & $ p>0.1$   & $p=0.012$ & $p=0.011$ & $p=0.036$ \\
Dooyoo			& $0.1881 $ & $ p<0.001$  & $0.1945 $ & $ p<0.001$  & $0.1856 $ & $ p<0.001$ & $p<0.05$ & $p<0.05$ & $p<0.05$ \\
{\small Rotten Tomatoes} & $0.0838 $ & $ p<0.01$  & $0.0929 $ & $ p<0.001$  & $0.0741 $ & $ p<0.001$ & $p<0.05$ & $p<0.05$ & $p<0.05$ \\
Redtube			& $-0.009 $ & $ p<0.001$  & $-0.008 $ & $ p<0.001$  & $-0.008 $ & $ p<0.001$ & $p=1$ & $p=1$ & $p=1$ \\
Pornhub			& $0.0332 $ & $ p<0.001$  & $0.0149 $ & $ p<0.001$  & $0.0113 $ & $ p<0.01$ & $p<0.05$ & $p<0.05$ & $p<0.05$ \\

\end{tabular}
\end{center}

\caption{Summary of results of the decoding study. Regression estimates and
significance levels for the right side coefficient $\hat{b}$ in Ordinary Least
Squares (OLS), robust regression (MM), and residualized estimates using
controls. Significance levels of one-tailed bootstrap, permutation, and Spearman tests with alternative hypothesis $b>0$, with maximum significance
level of $p<0.05$ in $10000$ samples. With the exception of Epinions,
BookCrossing, and Redtube, there is robust evidence that support the existence
of a QWERTY effect.   \label{tab:Decoding}}
\end{table*}

\begin{figure*}[ht]
\includegraphics[width=0.98\textwidth]{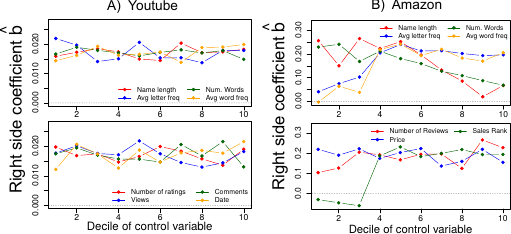}
\caption{Stratified analysis of the QWERTY effect in Youtube videos (A) and
Amazon products (B) controlling for linguistic measures (top) and contextual
properties (bottom). Each dot shows an estimate of the right side coefficient $\hat{b}$
for each decile divided by the control variable. The right side coefficient $\hat{b}$
is stable across all variables in Youtube. The right side coefficient in
Amazon  decreases for products with infrequent words and long names, and
disappears for products with very high sales.  \label{fig:AmazonControls}}
\end{figure*}

\textbf{Controlling for confounds:} The details of the models including
controls, i.e. after residualizing valence as a function of other variables,
are provided in Table \ref{tab:Decoding}. All residualized estimates are significant
and consistent with the rest of results, with the exception  of Epinions.
Thus, for the other 8 cases, the results of the other tests cannot be
explained as linear confounds with other variables, lending robust evidence
for the existence of the QWERTY effect.

To dig deeper in the role of linguistic properties of the names in  Youtube
and Amazon, we report the stratified analyses of the right side coefficient
$\hat{b}$ depending on each control variable.  Figure
\ref{fig:AmazonControls}A shows the right side coefficient $\hat{b}$ for
deciles varying control variables in Youtube.   A first observation yields a
simple conclusion, the QWERTY effect is independent of  the amounts of words
and letters in the title, as well as of their average frequencies. In
addition, the effect does not dramatically change with respect to other
controls, including the amount of views of a video, the amount of comments,
the total amount of likes and dislikes, nor on the date of the video. Thus we
find a strong robustness of the  effect in Youtube, even when accounting for
sample sizes when controlling for the amount of ratings of a video.

The case of Amazon is more interesting, as shown in Figure
\ref{fig:AmazonControls}B, the right side coefficient decreases for longer
product names. With respect to frequency, the effect becomes negligible for
products with letters and words of very low frequency. This analysis shows
that the QWERTY effect in Amazon weakens when infrequent or fabricated
language is used in product names, as well as with very long names. This
suggests that certain "tail of strangeness" elicits different responses in
raters that are not the same as for more recognizable names. With respect to
other controls, the right side coefficient $\hat{b}$ barely changed with price
or amount of reviews available, but showed a clear step function with respect
to sales rank. The effect was only present for products with  high rank
number, which means that they  were relatively low in the sales list. It is
important to note that the estimates for the best selling products are
slightly negative, showing that the effect might be reversed for superstar
products and some nonlinearities exist with respect to sales. This observation
does not contradict the residualized analysis of Table \ref{tab:Decoding}, as
the QWERTY effect is still positive in 7 deciles and it holds for the average
product when controlling for sales rank.

\section{Encoding study}
\label{sec:encoding}
\paragraph{Datasets}

We use four datasets of general product reviews for our encoding study,
using the text of reviews and their star ratings on a 1 to 5 scale. The
summary of the amount of reviews and aggregate rating at the review level are
provided in Table \ref{tab:data2}. For Amazon, we leverage the size of the dataset to avoid sampling
biases. From the more than 80 million reviews present in the dataset, we
randomly select at most one review per product and per user, to ensure that
extremely active users or extremely popular products do not bias our
statistics.

\begin{table}[hp]
\begin{center}
    \begin{tabular}{c|c|c|c|c}
	Dataset &   N reviews   & $\mean{r}$ & scale  \\ \hline
	Amazon & 971,026 & 4.0582 & 1-5 \\
	Yelp & 1,554,163 & 3.7412 & 1-5 \\
	Dooyoo & 523,997 &  4.0258 & 1-5 \\
	Epinions & 101,595 & 3.9768 & 1-5
	  \end{tabular}
\end{center}
  \caption{Summary of data used for the encoding study. \label{tab:data2}}
\end{table}

\paragraph{Methods}

We conduct an incremental regression model of  the rating of a review $V$ as a
function of text properties. We start from a null model that shows if there is
a relation between rating and review length  $V=a_l+b_l*N_l$, testing if
splitting the length data into $R$ and $L$ provides a better (adjusted)
estimate  when fitting a model with an interaction effect:
\begin{equation}
V=a_{RL} + b_{R}*R + b_{L}*L + b_{RL}*R*L
\end{equation}
This way we cope with the colinearity of $R$ and $L$.  If the QWERTY effect
manifests when encoding a review, the estimates should satisfy $b_{R}>0$ and
$b_{L}<0$, in a way such that their slopes do not change signs easily in the
range of possible values of $R$ and $L$. We test these hypotheses through
t-tests, and further verify them through permutation and bootstrapping as in
the Decoding Study.

\subsection{The QWERTY effect in encoding text}

Table \ref{tab:Encoding} shows the regression results for all four datasets.
For the cases of Amazon, Yelp, and Dooyoo, the model using $R$ and $L$ values
outperforms the length model, increasing the adjusted $R^2$. The estimates of
the coefficients in all cases follow the direction predicted by the QWERTY
effect: positive reviews tend to contain more right side letters and negative
reviews tend to contain more left side letters.  For the case of Epinions, the
signs of the estimates are in the predicted direction, but the estimates are
not significant. This is probably due to the limited size of the Epinions
dataset, and only more data or more powerful statistical methods could provide
a final answer.

\begin{figure*}[ht] \centering
\includegraphics[width=0.98\textwidth]{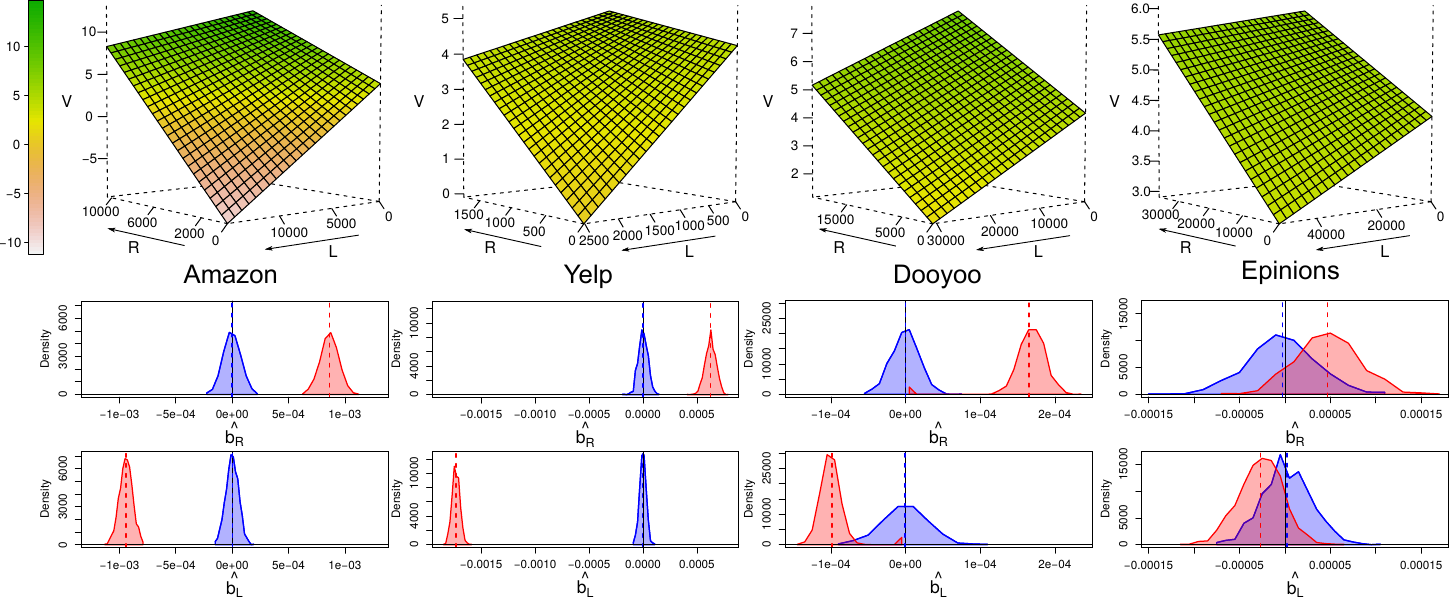}\\
\caption{Results of the interaction model for the four encoding datasets. Top:
The vertical dimension and color scale show the predicted value of the model,
versus the amount of right (R) and left (L) letters in the text. The surfaces
show that valence (V) increases with  the amount of right side letters (R) and
decreases with the amount of left side letters (L), even in the presence of
their interaction effect. Bottom: Results of permutation (blue) and bootstrap
(red) tests on the four datasets, confirming the conclusions of t-tests for parameters $b_R$ and $b_L$ in
Table \ref{tab:Encoding}. \label{fig:Encoding}} \end{figure*}

The interaction effect $b_{RL}$ is positive and significant in two cases. This
implies a nonlinearity in the role of the amounts of right and left side
letters which needs to be explored further. The top panel of Figure
\ref{fig:Encoding} shows the surfaces described by the regression results, for
the ranges of values of $R$ and $L$ present in the data. All cases present a
hyperbolic paraboloid shape, with slightly changing slopes in the ranges of
$R$ and $L$. A visual inspection shows how the nonlinear term $b_{RL}$ does
not change our conclusions: The slope on the $R$ axis is always positive and
only becomes moderated in the case of Yelp, but it does not turn negative in
the range of possible values. Similarly for $L$, the slope is always negative
and it only softens along the $R$ axis,  but it does not turn into a positive
slope. Thus, we conclude that, even in the presence of colinearities and
nonlinear effects, the QWERTY effect is present when encoding text to express
evaluative meaning.

\begin{table*}[bp]
\begin{center}
    \begin{tabular}{c|c|c|c|c|c|c|c}
    	&     	\multicolumn{2}{|c|}{Length model}    &  \multicolumn{4}{|c|}{$RL$ model}   & \\ \hline
Dataset	& $a_l$ & $b_l$ & $a_{RL}$ & $b_R$ & $b_L$ & $b_{RL}$ & $\Delta R^2_{adj}$ \\ \hline
Amazon &  4.096829  & -0.000123  & 4.110348 & 0.000863 & -0.000940 &0.0000001& 40.4\%  \\
Yelp &  4.172784  & -0.000458  & 4.263156 & 0.000625 & -0.001733 &0.000001& 8.2\%  \\
Dooyoo &  4.255313  & 0.000014  & 4.256541 & 0.000169 & -0.000101 &0.00 (ns)& 14.5\%  \\
Epinions &  4.274520  & 0.000007  &  4.278481 & 0.000048 (ns) & -0.000027 (ns) &0.00 (ns)& -5.1\%  \\
\end{tabular}
\end{center}
  \caption{Summary of results of the encoding study. All estimates are significant $(p<0.001)$ except those marked as (ns). Results show that, for three cases, the amounts of right side letters (R) and left side letters (L) explain additional (adjusted) variance than in the length model, and have significant estimates in the directions predicted by the QWERTY effect.\label{tab:Encoding}}
\end{table*}

The results shown in Table \ref{tab:Encoding} are robust to permutation and
bootstrapping tests. The lower panel of Figure \ref{fig:Encoding} shows the
density functions of the estimates of $b_R$ and $b_L$ after permuting the data
and in bootstrap samples. Permuted estimates for the first three datasets are
centered around $0$ and do not reach the mean of bootstrap estimates for three
of the datasets, and the their distribution of estimates in the bootstrap
samples is significantly far from $0$.

\section{Discussion}

In our studies, we find robust evidence for the QWERTY effect both at the
point of text interpretation (decoding) as well as at the point of text
creation (encoding). We explicitly avoided selecting on the dependent
variable by including a wide range of datasets in which the effect could be
tested. This way, we analyzed 15 datasets (11 in the decoding study and 4 in
the encoding study) from 11 different online media to obtain cross-platform insights. We found robust evidence for the existence of the
QWERTY effect  when reading texts (decoding) in 8 datasets, and we
learned that the effect might be reversed under particular conditions.
Furthermore, we found significant evidence during writing texts (encoding) in three datasets, showing that
positive reviews tend to contain more right side letters and negative reviews
more left side letters.\footnote{Our results can be replicated with the materials available at \url{https://github.com/dgarcia-eu/QWERTY_WWW}}

\paragraph{Limitations} Despite the general pattern of a positive right side
coefficient, two cases showed non-significant positive effects: Epinions and
BookCrossing. One explanation for this could be a lack of statistical power
in our strict methods, or the limited size of these two datasets. For the case
of BookCrossing, the effect could also be absent due to books existing before
widespread keyboard use. Another interesting exception appears in one of the
adult video communities, which was subject to possible sampling biases in the
listing of the sitemap. That dataset showed a strong cutoff on video dates and
might have had unpopular videos removed. If this were true, the negative
effect would be consistent with the interaction that we found in Amazon for
sales rank: The QWERTY effect weakens or even reverses when popularity or
marketing forces are in play. In any case, we must note that we lack a
testable conjecture that explains how the QWERTY effect might interact with
popularity and under which conditions.

The existence of the QWERTY effect in the datasets we have studied however
does not imply that it has to exist in every possible context, and its absence
in a community that discusses a particular type of content does not
constitute a rejection of the effect in general. Future work might
investigate the effect in additional web datasets on other types of products
or entities, such as beer \cite{Danescu2013} or wine \cite{Mcauley2013}, which
were not available to us during the course of our study.

We complemented previous experimental studies with
observational tests, and thus our contribution suffers the typical limitations
of observational, non-obtrusive studies. First, we did not have any degree of
control on the conditions of our tests and  we could not study if the same
content would be evaluated differently depending on its name. Natural
experiments are a promising avenue to combine experimental and observational
methods, but we ultimately need experiments in which, \emph{all things being
equal}, show a dependence between evaluation and right side letter
measurements. Second, besides our illustration of the RSR on books, we did not
reach timespans long enough to test the emergence of the QWERTY effect in a
longitudinal manner. This point is specially challenging, as we are dealing
with a property of language that cannot be simply reset in a group of
individuals.  Third, our observational methods only test the effect in an
aggregated manner, leaving open the question of individual idiographic
differences depending on demographic conditions like age or gender, on factors
of human-computer interaction like handedness or other layouts, and on
psychological factors like personality or locus of control.

Our work focused on English-speaking online media and filtered out non-English
content. We must highlight that this contains the assumption that the QWERTY
keyboard layout is widely used in those communities, which might not be the
case when non-native English speakers communicate online. Further research
might also consider other languages and keyboard layouts, but this might not
be straightforward, as languages coexist in some online media. It is also
important to note that the normative nature of language can cancel out
minority effects like left-handed users or other language keyboards, in line
with the findings that the effect can also appear in experiments with left-handed
users and other layouts \cite{Casasanto2014}. While we include a wide variety
of linguistic and contextual controls, we did not account for more complex
effects such as alphabetic ordering effects or finger alternations
\cite{Jasmin2012}. In addition, language is more than just letters, and
nonlinear hierarchical interactions should be tested at the level of words and
phrases.

While our work tests the existence of the QWERTY effect, we do not evaluate
its predictive power. Our regression approach allows us to quantify the
strength of the coupling between evaluative content and the amounts of right
and left side letters in item names and review texts.  As suggested by the
subconscious aspect of the effect, the strength of the Right Side Ratio was
very small, in the line of other influential results in psycholinguistics that
reveal variations of term frequencies below 1\% \cite{Golder2011,Kramer2014}.
A small change in emotional language can also have
large implications, for example driving social factors of emotions beyond
critical thresholds that produce macroscopic phenomena
\cite{Schweitzer2010,Alvarez2015}.

\paragraph{Dactilar onomatopoeias} The QWERTY effect is a special case of
\emph{onomatopoeia}, a symbol class that violates the general arbitrarity of
the relationship between signs and meanings \cite{DeSaussure1966}. This kind
of relationships are not new and are well documented in linguistics, in
general with respect of how words phonetically imitate the sounds or objects
they signify. For example, low voice tones are used to refer to large objects
(e.g. "huge") and high pitches to small objects (e.g. "tiny"), resembling how
large cavities produce lower sounds (e.g. contrabass vs violin). Since the
organs used to articulate words online are our fingers, we can talk about
\emph{dactilar onomatopoeias} that connect hands and fingers with meanings.
Pronouncing words with the vocal tract does not have any precise spatial
orientation, but using a keyboard to communicate introduces the asymmetry of
the keyboard layout in the communication process.

\section{Conclusions}

To the best of our knowledge, this study provides first evidence of the extent of the QWERTY effect on the web. In our
datasets we confirm that products with more right side letters and fewer left
side letters have higher average ratings, as suggested by \cite{Jasmin2012}.
As an application to marketing, our results support the concept that \emph{the
"right" name} might provide certain yet limited advantages. Furthermore, we
show that this extends to movie titles and Youtube videos, where the presence
of the effect in the ratio of likes was particularly robust. Our exploration
of confounds and interactions showed how the effect diminished for Amazon
products with infrequent words and very high sales. We also found an
interesting exception of an inverse effect that calls for further research to
understand how contextual properties can influence hand to meaning
associations. Finally, we tested the effect during encoding through the
text of product reviews and their star ratings, finding first evidence that positive meanings are written with more right side letters.

Our work has implications for wider research in psychology and computational social
science. The existence of the effect is an example of a combination of
collective social factors and embodied emotions within the hyperlens model of
emotions \cite{Kappas2013}. Our results can also be understood as an example
of lateralization, not only in the metaphoric sense but also as a
psychological phenomenon. With respect to online interaction, the QWERTY
effect is one example of social components in the expression of sentiment, in
line with assortativity of happiness \cite{Bollen2011} and emotion-dependent
cascading behavior \cite{Alvarez2015}. 

The right and left side letter quantities that we measured are promising with
respect to tasks that involve semantic annotation or sentiment analysis.
Future sentiment analysis tools might explore whether adding the Right Side
Ratio as a variable improves prediction quality. In addition, it will be
interesting to explore whether and how the QWERTY effect might interact with
popularity and spreading processes, testing if messages with more right side
letters are shared more frequently or trigger longer discussions.\\

\textbf{Acknowledgments:} 
DG was funded by the Swiss National Science Foundation (CR21I1\_146499/1). We
thank Simon Schweighofer and Cristina Soriano for useful discussions.

\clearpage
\small
\raggedright
\sloppy

\end{document}